\documentclass[a4paper,11pt]{article}
\pdfoutput=1 

\usepackage{jheppub} 

\newcommand{\beq}{\begin{equation}}
\newcommand{\eeq}{\end{equation}}
\newcommand{\rn}{Reissner–Nordstr\"om}

\title{Islands and Page curves of {\rn} black holes}
\author[a,*]{Xuanhua Wang,}
\author[b,c,*]{Ran Li,\note[*]{Equal contributions}}
\author[a,b,1]{Jin Wang\note{Corresponding author}}
\affiliation[a]{Department of Physics and Astronomy, Stony Brook University, Stony Brook, NY 11794, USA,}
\affiliation[b]{Department of Chemistry, State University of New York at Stony Brook, Stony Brook, NY 11794, USA}
\affiliation[c]{Department of Physics, Henan Normal University,
 Xinxiang 453007, China}
\emailAdd{xuanhua.wang@stonybrook.edu}
\emailAdd{liran@htu.edu.cn}
\emailAdd{jin.wang.1@stonybrook.edu}

\abstract{We apply the recently proposed quantum extremal surface construction to calculate the Page curve of the eternal Reissner-Nordstr\"om black holes in four dimensions ignoring the backreaction and the greybody factor. Without the island, the entropy of Hawking radiation grows linearly with time, which results in the information paradox for the eternal black holes. By extremizing the generalized entropy that allows the contributions from the island, we find that the island extends to the outside the horizon of the Reissner-Nordstr\"om black hole. When taking the effect of the islands into account, it is shown that the entanglement entropy of Hawking radiation at late times for a given region far from the black hole horizon reproduces the Bekenstein-Hawking entropy of the Reissner-Nordstr\"om black hole with an additional term representing the effect of the matter fields. The result is consistent with the finiteness of the entanglement entropy for the radiation from an eternal black hole. This facilitates to address the black hole information paradox issue in the current case under the above-mentioned approximations.}

\begin{document} 
\maketitle
\flushbottom

\section{Introduction and motivation}

The information issue of black holes is a fundamental problem in several most important fields of physics--quantum mechanics, thermodynamics and the theory of general relativity, and is essential for our understanding of quantum gravity \cite{Hawking:1974sw,Page:1993wv,Wald:1975kc,Parker:1975jm}. Recently, tremendous progress has been made to provide a quantum description of the information conservation in the process of black hole evaporation \cite{Almheiri:2019hni,Almheiri:2019yqk,Almheiri:2019psf,Penington:2019npb,Engelhardt:2014gca,Almheiri:2019qdq}. This was done without the recourse to a complete understanding of the quantum dynamics of black holes, which seems to necessarily involve the understanding of quantum gravity. 

The origin of the information paradox dates back to decades ago. In 1975, Hawking proposed that the information falling into the black hole would disappear after the evaporation of the black hole \cite{Hawking:1974sw,Parker:1975jm}. However, this proposed process violates unitarity, which is one of the foundations of quantum mechanics. According to the principle of unitarity, the evaporation of a black hole from a pure state with zero entropy has to end with the pure-state quantum gas of radiation instead of mixed-state thermal gas which has a large entropy. This argument incorporating the expected initial and final state behaviors is represented graphically through the Page curve \cite{Page:1993wv,Page:1993df}. Page showed that for a quantum system such as a black hole to evolve unitarily from an initial pure state, the entanglement entropy increases linearly with time in the initial period of the evaporation and gradually decays to zero after the black hole radiates the majority of its energy. Many other proposals were suggested to resolve the information paradox. Some representative ones and their pros and cons were discussed in Ref.~\cite{Giddings:1992fp}. One approach suggested to include backreaction leading to a final pure state. But it appears to imply that either all of the information has been extracted by the time the falling matter crosses the horizon or that information escapes acausally from the black hole \cite{Page:1979tc}. Another proposal suggested information release at the end of black hole evaporation at the Planck scale. But this proposal requires the remaining Planck scale energy to carry off arbitrarily amount of information which would violate the Bekenstein bound \cite{Hawking:1982dj,Bekenstein:1980jp,Harlow:2014yka}. A different proposal suggested the Planck scale remnant after the evaporation. But the remnant is intrinsically unstable \cite{Banks:1983by}. A proposal on including baby universes as a source of information loss was suggested. But later studies showed that wormholes only change the coupling without violation of unitarity \cite{Giddings:1988cx}. There was also a proposal on a previously unexpected mechanism of information release. But this suggestion seems to require the violation of causality on the horizon \cite{Giddings:1992hh}. In 1999, Parikh and Wilczek proposed to address the information paradox issue by including the higher order non-thermal effect in the radiation to allow information to leak out from the black hole \cite{Parikh:1999mf}. However, this effect is negligible for massive black holes and not able to compensate for the information loss in this case. 

Whether black hole dynamics preserve unitarity remained a conundrum until present. One of the breakthrough ideas was made by the discovery of AdS/CFT correspondence \cite{Maldacena:1997re}. The duality is a mathematical realization of the proposed idea of the black hole complementarity \cite{Susskind:1993if}, and provides a strong evidence for the conservation of information as the black hole in the anti-de Sitter space (AdS) can be mapped to the boundary CFT. Therefore, the evaporation of the black hole has a dual unitary description using the boundary CFT. If this argument is true, the evaporation of black holes should roughly follow the Page curve. However, the quest to obtain the Page curve remains unsuccessful until very recently. Apart from that, the unitary process was shown to generate a ``firewall" (AMPS firewall) on the black hole horizon which is at odds with the ``no drama" principle of the general relativity \cite{Almheiri:2012rt}.  For eternal black hole, similar questions on the information paradox can be addressed. For a unitary evolution, the corresponding Page curve is expected to reach a bounded value which is the Bekenstein-Hawking entropy of the black hole. The amount of radiation for a eternal black hole is infinite at the ``end stage" or the late times of the evaporation. Thus, a thermal spectrum of radiation would produce an infinite amount of entropy. This is contradictory to the unitarity which dictates the maximal entropy produced by the black hole to be the Bekenstein-Hawking entropy. A resolution to all the issues related to the black hole information paradox has been long-yearned.

The Page curve of Hawking radiation was recently calculated by using the semi-classical method for two-dimensional black holes in asymptotically AdS spacetimes in Jackiw-Teitelboim (JT) gravity \cite{Almheiri:2019qdq,Almheiri:2019yqk}. Most of the studies on the the black hole information problem have been concentrated to two dimensional gravity where exact solutions incorporating the backreaction of the radiation are possible under semi-classical approximations \cite{Rozali:2019day,Chen:2019uhq,Chen:2020jvn,Chen:2020uac,Chen:2020hmv,Hernandez:2020nem,Akers:2019lzs,Liu:2020gnp,Marolf:2020xie,Verlinde:2020upt,Chen:2020wiq,Gautason:2020tmk,Anegawa:2020ezn}. In the two dimensional systems, islands appear at the later stage of the black evaporation, which is in the entanglement wedge of the radiation, such that the Bekenstein bound of the entanglement entropy is preserved. For a review see Ref.~\cite{Almheiri:2019qdq}. However, whether this island construction can be extended to and resolve the information issue of all black hole solutions still remains to be verified. For higher dimensional or ``realistic" black holes in four dimensional asymptotic spacetime, the resolution of the information paradox is much less studied due to the difficulty in calculating the entanglement entropy and analysing the dual conformal field theory (CFT). It is argued in \cite{Penington:2019npb} that the islands should exist in the higher dimensional black hole spacetimes and the unitary Page curves can also be reproduced if taking the island's effect into account. Recently, some interesting phenomenological studies of the island structure and the Page curves in four dimensional Schwarzschild and dilaton black hole were performed in Ref.~\cite{Hashimoto:2020cas,Karananas:2020fwx}. Some other studies of different models in three dimensions and higher dimensions can be found in refs.~\cite{Alishahiha:2020qza,Caceres:2020jcn,Hashimoto:2020cas,Gautason:2020tmk,Karananas:2020fwx,Almheiri:2019psy,Ling:2020laa,Krishnan:2020oun,Krishnan:2020fer,Geng:2020fxl,Geng:2020qvw,Bak:2020enw,Hartman:2020khs,Balasubramanian:2020coy,Balasubramanian:2020xqf,Sybesma:2020fxg,Manu:2020tty,Choudhury:2020hil,Bhattacharya:2020ymw,Deng:2020ent,Verheijden:2021yrb,Li:2020ceg}.

Our present understanding of the entropy of quantum systems coupled to gravity does not necessarily requires holography and AdS space \cite{Almheiri:2019qdq}, it is nevertheless an essential tool in the development of the entropy of gravitational systems. The groundbreaking work of Ryu and Takayanag (RT) using AdS/CFT correspondence connects the entanglement entropy of the boundary region to the area of the minimal surface in the bulk space \cite{Ryu:2006bv}. Later works generalized the RT surface to the quantum extremal surface, in which the generalized entropy includes all the quantum corrections of the bulk fields \cite{Engelhardt:2014gca,Faulkner:2013ana,Barrella:2013wja,Lewkowycz:2013nqa,Hubeny:2007xt,Penington:2019npb}. It is shown that by applying the extremal surface technique, islands appear at the later stage of black hole evaporation process, and that the entropy of Hawking radiation obeys the Page curve assuming the unitary \cite{Almheiri:2019qdq}. Furthermore, the island formula for the fine-grained entropy of the Hawking radiation is proposed to be \cite{Faulkner:2013ana,Engelhardt:2014gca,Wall:2012uf,Akers:2019lzs}
\begin{equation}
\label{GEformula}
 S(R) 
 = 
 \min \left\{\mathrm{ext}\left[
 \frac{\mathrm{Area}(\partial I)}{4G_{\rm N}} 
 + S_{\rm matter}(R\cup I)
 \right]\right\} \, 
\end{equation}
where $R$ is the radiation, $I$ is the island, and $S_{\rm matter}$ is the entropy of quantum fields. It is shown that the island formula can be derived without holography from the Euclidean path integral by using the gravitational replica method. The presence of replica wormholes as the saddle points in the Euclidean path integral leads to the island formula not only for the eternal black holes but also for the evaporating black holes \cite{Almheiri:2019qdq,Penington:2019kki,Hartman:2020swn,Goto:2020wnk}.

The black hole information paradox has been addressed mainly in two-dimensional gravity models. It is equally important, if not more, to resolve the paradox in our real universe, which is four-dimensional and reaches Minkowski space asymptotically. The full solutions including backreaction in higher dimensions are highly nontrivial due to the nonlinearity of Einstein's equations. The static solutions ignoring the backreaction, on the other hand, are possible to handle after taking several approximations and reveal some of the most important properties of the island configurations. In the classical general relativity and cosmology, there are a few important 4D vacuum solutions to Einstein's theory of relativity that deserve particular attention. In this article, we will address the information paradox issue for the four dimensional charged black hole solution in the asymptotically flat spacetime and study the island structure under the s-wave approximation. In this study, we construct the Page curve for the four dimensional eternal Reissner–Nordstr\"om black holes in the asymptotically flat spacetime and show that the entanglement island in this case saves the entropy of the radiation from exploding at the late times. This quantitatively resolves the information paradox for the {\rn} black hole. 

In this work, we will apply the method of quantum extremal surface to study the entropy of Hawking radiation and the corresponding Page curve of the {\rn} spacetime in four dimensions. The action is given by the Einstein-Maxwell action 
\begin{align}
 I &= \frac{1}{16\pi G_{\rm N}} \int_{\mathcal M} d^4 x \sqrt{-g} \,\left( R 
 -\frac{1}{4}F^{\mu\nu}F_{\mu\nu}\right) + I_{\rm matter} \, ,
\label{EinsteinHilbert}
\end{align}
where $G_{\rm N}$ is the Newton constant, and $I_{\rm matter}$ is the action of the matter fields. If the matter fields are described by the CFT and are in the vacuum states, the vacuum solution to the Einstein-Maxwell action will not be affected by the matter fields. It is straightforward to generalize the analysis to gravity with higher curvature terms, but we focus only on the dominant contributions.

The metric of the Reissner–Nordstr\"om black holes is given by 
\beq ds^2=-\left(1-\frac{2M}{r}+\frac{Q^2}{r^2}\right)dt^2+\left(1-\frac{2M}{r}+\frac{Q^2}{r^2}\right)^{-1}dr^2+r^2d\Omega^2\,,
\eeq
where we have set the Newton's constant and the Coulomb constant equal to 1, i.e. $G_N=K=1$. These physical constants can be easily restored if necessary. {\rn}  spacetime is one of the most important vacuum solutions of the Einstein's field equation representing a charged black hole in the 4D asymptotically Minkowski space. One of the distinctions of the {\rn} black hole from the Schwarzschild black hole spacetime is the appearance of two horizons (event horizon and causal horizon) even though the inner causal horizon is believed to be unstable under small perturbations due to the mass inflation phenomenon. 

In the present work, we only consider the non-extremal black holes. The radius of the horizons are $r_{\pm}=M\pm\sqrt{M^2-Q^2}$ and the surface gravity at the horizons is given by $\kappa_{\pm}=\frac{r_{\pm}-r_{\mp}}{2r_{\pm}^2}$. The Hawking temperature of the {\rn} black hole is given as follows, 
\beq T_{RN}=\frac{\kappa_+}{2\pi}\,. \eeq

This paper is arranged as follows. In section 2, we present an approximate method to compute the entanglement entropy for quantum fields in four dimensions. In section 3, the entropy of Hawking radiation is computed without island and the corresponding information paradox for the eternal {\rn} black hole is sharpened. In section 4, we analyze the generalized entropy of Hawking radiation and reproduce the unitary Page curve when taking the effect of islands into account. Based on these results, we also discuss the Page time and scrambling time for the {\rn} black holes in section 4.3. The discussion and conclusion are presented in the last section.


\section{The entanglement entropy: general approach}
\label{sec:woisland}

In the following sections, we carry out the calculation of the entanglement entropy in the four dimensional Reissner–Nordstr\"om geometry without/with involving the islands. The entanglement entropy for a general four dimensional spacetime is not known. However, the Hawking radiation observed from a distant observer can be properly described by the s-wave approximation. For {\rn} black holes, the Hawking radiation and Schwinger effect both act as the emission channels of charged pairs. We only consider the neutral uncharged radiation from the black hole and neglect the Schwinger effect. In order for the Hawking process to dominate over the Schwinger process, we require that the energy of the black hole is much larger than its charge. In addition, we assume that the black hole is macroscopic so that the backreaction is ignored and the central charge $c$ satisfies $1\ll c \ll M$. Under these assumptions, the dynamics of the radiation is effectively a two dimensional CFT described semi-classically. We ignore the grey-body factor and apply the analysis of the 2-dimensional CFT to obtain the approximate entanglement entropy in the curved 4-dimensional spacetime.

For the one dimensional quantum many-body systems at critiality (i.e. CFT in two-dimensions), it is known that the entanglement entropy neglecting the UV-divergent part (or Plank scale physics) is given as follows, 
\beq S_A=\frac{c}{3}\cdot \log(\frac{L}{\pi \epsilon}\sin(\frac{\pi l }{L}))\simeq \frac{c}{3}\cdot \log l\,,
\label{cft}
\eeq
where $l$ and $L$ are the lengths of the subsystem $A$ and the total system,  $\epsilon$ is the UV cutoff, and $c$ is the central charge of the CFT. We have assumed that $l\ll L$ and kept only the finite part.

As is shown by Ryu and Takayanagi, the entanglement entropy in the boundary (d+1)-dimensional CFT has a dual description in the bulk. It follows a simple area law when mapped into the bulk, i.e.
\beq S_A=\frac{\mathcal{A}}{4G_N^{d+2}}\,, \label{rt}\eeq
where $\mathcal{A}$ is the area of the the d-dimensional static minimal surface in the AdS$_{d+2}$ \cite{Ryu:2006bv}. For a two-dimensional CFT, it is just the length of the minimal curve in the bulk. This formula is applied when no island is formed.

For two dimensional systems with multiple disjoint intervals $A=\{x|\ x\in [r_1,s_1]\cup[r_2,s_2]\cup...\cup[r_N,s_N]\}$, the generic formula for the entanglement entropy derived from the Ryu-Takayanagi formula is given as 
\beq S_A=\frac{\sum_{i,j}L_{r_j,s_i}-\sum_{i<j}L_{r_j,r_i}-\sum_{i<j}L_{s_j,s_i}}{4G_N} \,.
\label{generalrt}
\eeq
where $L_{r_j,s_i}$ is the minimal surface in the bulk with the boundary points $[r_j,s_i]$ and $G_N$ is the Newton's constant in 3 dimensions. The above equations, Eq.~\eqref{cft},~\eqref{rt} and \eqref{generalrt} allow us to compute the entanglement entropy for disjoint union of intervals. This general RT formula will be useful when one or more islands appear. For the simplicity of the study, we restrict to the case that either has no island or only one island. In principle, any possible number or shape of islands is allowed but this also makes the analysis extremely tedious. We will show that it is sufficient to resolve the information problem of the black hole and gives the sensible Page curve restricting the calculation to one island. 

With the approach to calculate the entropy, we apply the quantum extremal surface with the islands neglecting the backreaction of the radiation on the black hole metric. First, we use the explicit expression for the generalized entropy,
\beq 
S_{\rm gen}= \frac{\mathrm{Area}(\partial I)}{4G_{\rm N}} 
 + S_{\rm matter}(R\cup I) \,,
 \eeq
and then extremize it with respect to time and spacial coordinates of the island. If no saddle point is found, then we claim that no island will form in that case. Otherwise, we include the configurations of the islands in the generalized entropy and take the minimal values of all such saddle points. If the island configuration can resolve the information paradox, we expect that the entropy at late times reaches a finite value which is bounded by the Bekenstein-Hawking entropy. Otherwise, the information will not be conserved and the information paradox will remain to be an conflicting issue of gravity and quantum mechanics for {\rn} black holes.

\section{The diverging entanglement entropy without island}\label{sec:noislands}

\begin{figure}[t]
\centering
	\includegraphics[width=75mm]{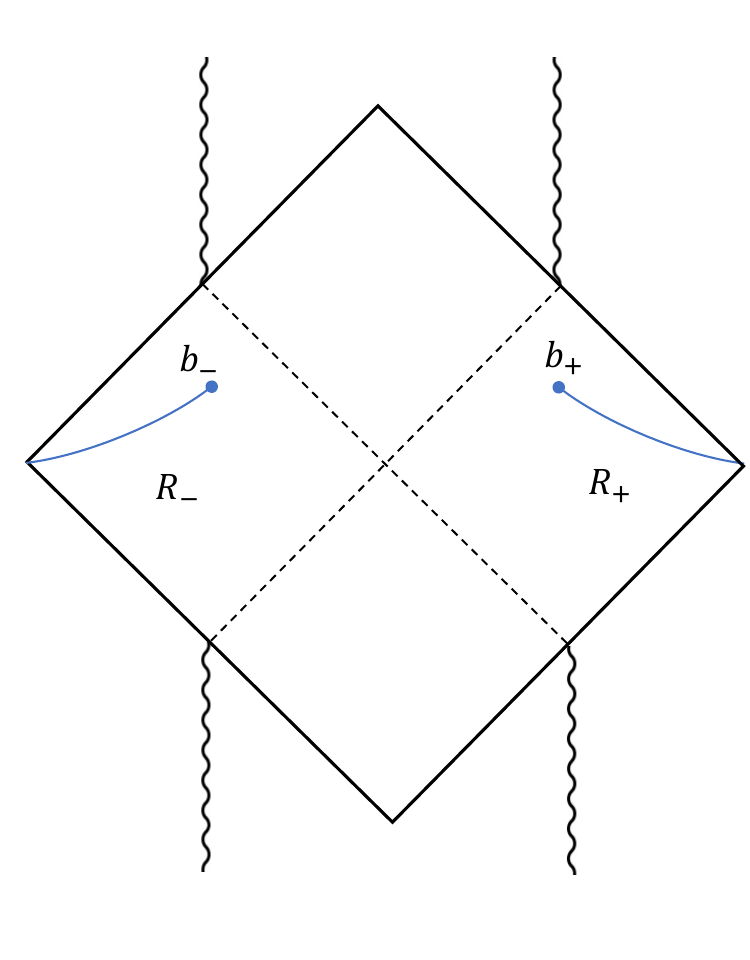}
	\caption{Penrose diagram for the eternal {\rn} black hole without islands. Hawking radiation is assumed to lie on the region $R_+$ and $R_-$. $b_{\pm}$ are the boundary surfaces of $R_+$ and $R_-$, respectively.}
	\label{fig:penrose}
\end{figure}

In this section, we will calculate the entanglement entropy of the radiation at late times without considering the contributions of the islands and sharpen the information paradox for the {\rn} black holes. In Fig.~\ref{fig:penrose}, the region of space where the radiation is "collected" or counted is the blue lines with cutoff points $b_{\pm}$. We assume that the region is far from the black hole and the radiation can be approximated as CFT in flat space. In the absence of the island, we have only two endpoints for the entanglement region of the radiation. They are the boundary points $b_{\pm}$ of the region $R_+$ at the right and the one on the left $R_-$ (see Fig.~\ref{fig:penrose}). 

Assuming that no island is formed at the late times of the evaporation process, we refer the entanglement entropy to Eq.~\eqref{cft}. If we assume the whole system is in the pure state at $t=0$,  the entropy of the CFT in region outside [b-; b+] is the same as the entropy within the interval. Therefore, we have
\begin{equation}
 S_{\text{R}} =  \frac{c}{3} \log d(b_+,b_-)\;,
\end{equation}
where $b_+$ and $b_-$ stand for the boundaries of the entanglement regions in the right and the left wedges of the Reissner–Nordstr\"om geometry and $d(b_+,b_-)$ is the distance between the points $b_+$ and $b_-$. Here $(t,r) = (t_b,b)$ for $b_+$, and $(t,r) = (-t_b + i {\beta}/{2},b)$ for $b_-$, respectively. Here, $\beta=2\pi/\kappa_+$ is the inverse of the Hawking temperature and $\kappa_+$ is the surface gravity at the outer horizon.

Though the existence of the inner horizon in the Reissner–Nordstr\"om black hole will result in the space not admitting a Cauchy surface like the case of the Schwarzschild spacetime, the inner horizon in the Reissner–Nordstr\"om black hole is unstable under small perturbations \cite{waldrn2020}. On the other hand, allowing the inner horizon will cause many other issues such as the violation of strong cosmic censorship, therefore, the effect of the inner horizon is ignored in this study. We will see latter that the assumption is justified. Under this assumption, we refer to the Kruskal-Szekeres-like coordinate in the Reissner–Nordstr\"om spacetime which is given by \cite{pradhan10} 
\begin{equation}
 ds^2 = - \frac{r_+r_-}{r^2\kappa_+^2}(\frac{r_-}{r-r_-})^{\frac{\kappa_+}{\kappa_-}-1}e^{-2\kappa_+r}dUdV + r^2 d \Omega^2 \ ,
 \label{dUdV} 
\end{equation}
where we have defined the coordinate variables as
\begin{align}
& r_* = r+ \frac{r_+^2}{r_+-r_-}\log|r-r_+|-\frac{r_-^2}{r_+-r_-}\log|r-r_-| \ , \label{rstar} \\
 & U \equiv - e^{-\kappa_+(t-r_\ast)} \, ,\ \rm{and} \quad
 V \equiv e^{\kappa_+(t+r_\ast)}\, .
\end{align}
For simplicity, we denote the conformal factor of the Reissner–Nordstr\"om geometry as $f^2(r)$, i.e.
\begin{align}
f^2(r) = \frac{r_+r_-}{r^2\kappa_+^2}(\frac{r_-}{r-r_-})^{\frac{\kappa_+}{\kappa_-}-1}e^{-2\kappa_+r} \ . 
\end{align}
Then, the metric after the conformal map is simply as follows,
\beq 
ds^2 = - f^2(r)dUdV + r^2 d \Omega^2 \,.
\eeq
Following the conformal mapping, the matter part of the entanglement entropy in
the Reissner–Nordstr\"om geometry is
\begin{equation}
 S_{\text{R}} = \frac{c}{6} \log \left[f(b_+)f(b_-)\left(U(b_-) - U(b_+)\right)\left(V(b_+) - V(b_-)\right)\right] \, .
\end{equation}
The total entanglement entropy by applying formula Eq.~\eqref{cft} is calculated as 
\begin{equation}
 S = 
\frac{c}{6} \log \left[4 f(b)^2e^{2\kappa_+r_*(b)} \cosh^2\kappa_+t\right]\,,\label{noislands}
\end{equation}
where 
\beq r_*(b) = b+ \frac{r_+^2}{r_+-r_-}\log|b-r_+|-\frac{r_-^2}{r_+-r_-}\log|b-r_-|\,.\eeq 
This result, i.e. Eq.~\eqref{noislands} gives the black hole fine-grained entropy as a function of time when no island is considered. We notice that this entropy blows up as $t\rightarrow \infty$,
\beq
S \sim  \frac{c}{6}\log(4\cosh^2\kappa_+t)\sim \frac{c}{3} \kappa_+t\,. 
\eeq

Therefore, without island the information does not leak out of the black hole and the entanglement entropy increases linearly with time. No Page time shows up in this calculation and the entropy of the radiation will eventually be infinitely larger than the Bekenstein-Hawking entropy of the black hole. Assuming that the eternal black is sustained by feeding it on pure-state quanta, the total von Neumann entropy of the black hole does not change and the entanglement entropy of radiation is at most double the Bekenstein-Hawking entropy (left and right regions in the conformal diagram). Therefore, there is clearly a paradox from the above calculation. We shall see that the island construction will resolve this issue and predict the correct Page curve for eternal Reissner–Nordstr\"om black holes in the next section.

\section{The entanglement entropy with island}
\label{sec:wisland}

\begin{figure}[t]
\centering
	\includegraphics[width=80mm]{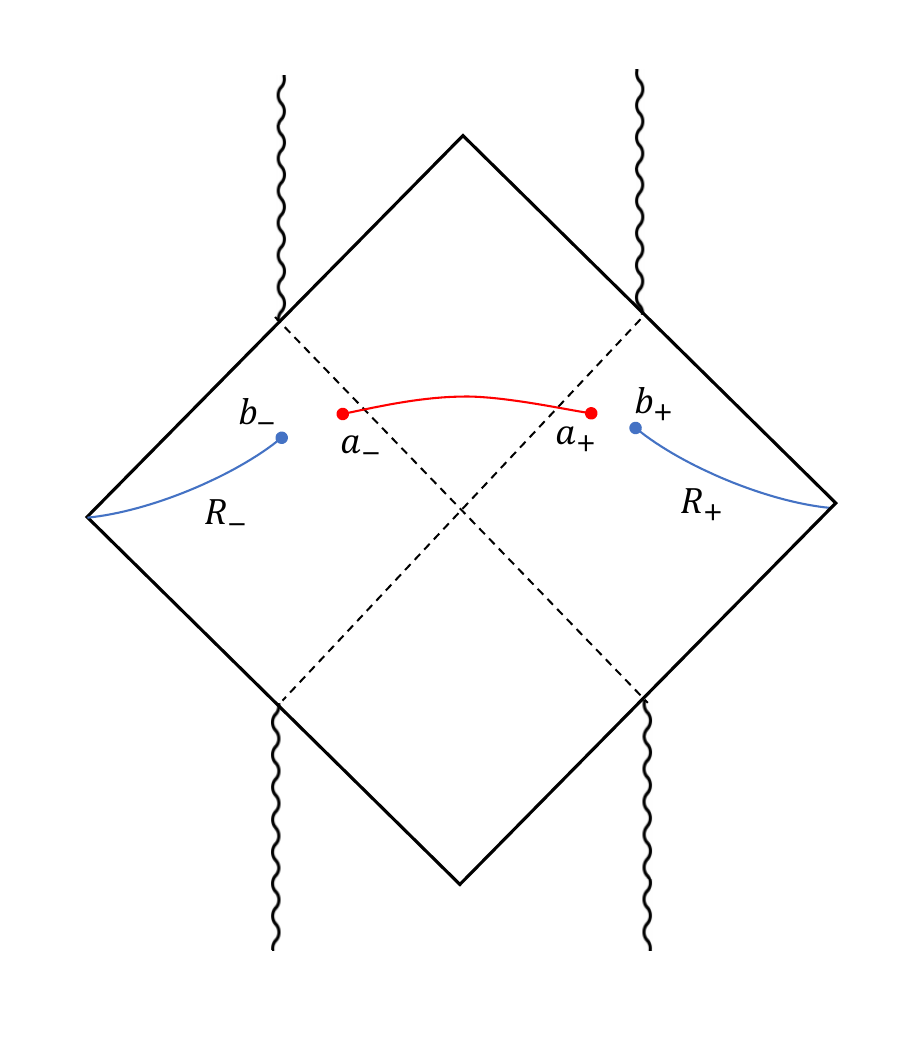}
	\caption{Penrose diagram for the eternal {\rn} black hole with the assumption of island. The islands extend to the outside of the horizons of the black holes. The boundaries of islands are located at $a_+$ and $a_-$. The points $b_\pm$ are the boundaries of the left and right radiation regions $R_-$ and $R_+$. }
	\label{fig:penroseisland}
\end{figure}

In this section we will introduce island construction and calculate the entanglement entropy with a single island. For eternal Reissner–Nordstr\"om black hole, we are only interested in the long-time limit of the entanglement entropy, i.e. the radiation at late times since it is when the entropy becomes an issue. We consider the case when the boundary $r=b$ of the entanglement region $R$ is far away from the horizon, $b \gg r_{\rm h}$. Besides, we assume that the s-wave approximation is valid and use the matter entropy formula for calculating the total entropy. 

The expression of the entanglement entropy for the conformal matter is inferred by Eq.~\eqref{generalrt},
\begin{equation}
 S_{\text{matter}} = \frac{c}{3} \log \frac{d(a_+,a_-) d(b_+,b_-) d(a_+,b_+) d(a_-,b_-)}{d(a_+,b_-) d(a_-,b_+)} \ .
\end{equation}
Using the Kruskal coordinates given in Sec.~\ref{sec:woisland}, the generalized entropy is the semi-classical fine-grained entropy in Eq.~\eqref{generalrt} plus the area of the quantum extremal surface as is given by 
\begin{align}
 S_{\mathrm{gen}} &= \frac{2\pi a^2}{G_{\rm N}} + \frac{c}{6} \log \left[2^4 f^2(a)f^2(b)e^{2\kappa_+(r_\ast(a)+r_\ast(b))}\cosh^2(\kappa_+t_a)\cosh^2(\kappa_+t_b)\right] 
\notag\\
 &\quad
 + \frac{c}{3} \log \left[\frac{\cosh\left(\kappa_+(r_*(a)-r_*(b))\right)-\cosh\left(\kappa_+(t_a-t_b)\right)}
 {\cosh\left(\kappa_+(r_*(a)-r_*(b))\right) + \cosh\left(\kappa_+(t_a+t_b)\right)}\right] \ , 
\label{islands}
\end{align}
where $r_\ast(a)$ and $r_\ast(b)$ are defined in Eq.~\eqref{rstar} and $f^2(x)$ is the conformal factor of the metric,
\begin{align}
f^2(r) = \frac{r_+r_-}{r^2\kappa_+^2}(\frac{r_-}{r-r_-})^{\frac{\kappa_+}{\kappa_-}-1}e^{-2\kappa_+r} \ . 
\end{align}
and the $r_\ast(a)$ is defined to be
\begin{align}
& r_*(a) = a+ \frac{r_+^2}{r_+-r_-}\log|a-r_+|-\frac{r_-^2}{r_+-r_-}\log|a-r_-| \, .
\end{align}

The first term in Eq.~\eqref{islands} corresponds to the area of the island, and the second and third term correspond to the entropy of the matter outside the cutoff surface and inside the island $R\cup I$. To get the entanglement entropy we still need to extremize the formula of the generalized entropy Eq.~\eqref{islands} over all possible Cauchy surfaces and take the minimal value. However, before doing so we can already get some information from this formula. At the first glance, this formula also appears to explode as the time goes to infinity. However, one should keep in mind that we have introduced free parameters into the formula. When we extremize Eq.~\eqref{islands}, we can hope that this free parameter will result in a bounded answer as time goes to the infinity. We will see later that this is the correct guess.

\subsection{Early times}
In the following, we study the early and late times behavior of the island the  entropy. The structure of the island can be analysed as follows. At early times, the entanglement entropy of the black hole with the radiation is small and thus the extremal surface that can globally minimize the generalized entropy has to lie deep inside the black hole if it exists at all. We assume that $t_a,\ t_b \ll r_+$ and we pick the cutting surface far away from the horizon $b\gg r_+$. Then the last term in the generalized entropy Eq.~\eqref{islands} can be properly ignored and the generalized entropy is approximately as follows, 
\begin{align}
    S_{\rm gen} &\simeq \frac{2\pi a^2}{G_{\rm N}} + \frac{c}{6} \log \left[2^4 f^2(a)f^2(b)e^{2\kappa_+(r_\ast(a)+r_\ast(b))}\cosh^2(\kappa_+t_a)\cosh^2(\kappa_+t_b)\right] \notag\\
    &\simeq \frac{2\pi a^2}{G_N}+\frac{c}{3}(\kappa r_*(a)+\log f(a))+\text{functions of ($t_b$, $r_*(b)$) which will be omitted} \notag\\
    & \simeq \frac{2\pi a^2}{G_N}+\frac{c\kappa_+}{3}\left(\frac{r_+^2}{r_+-r_-}\log |a-r_+|+(\frac{r_-^2+r_+^2}{2r_+^2}-\frac{r_-^2}{r_+-r_-})\log |a-r_-|-\log a\right)\,.
    \label{early}
\end{align} 
Extremize this function over $a$ under the approximation that $t_a,\ t_b\ll r_*(b)-r_*(a)$, we find that \beq a\approx \sqrt{\frac{c}{12\pi}}\cdot l_P \eeq where $l_P$ is the Plank length. Naively we get a Plank scale island inside the black hole which can store the minimal amount of information. However, the upper cutoff length of the above approach is far above Plank length and we have thrown away all Plank scale physics. Besides we cannot draw any Cauchy surface into the inner horizon and it is not covered by the metric we adopted. This controversy really means that in the regime where our analysis applies, there does {\it not} exist a nonvanishing quantum extremal surface that minimize the generalized entropy. On the other hand, we pick the minimum of two configurations with or without the island. This can be verified by comparing the entropy achieved above with the entropy calculated without island Eq.~\eqref{noislands}, \begin{equation}
 S = \frac{c}{6} \log \left[4 f(b)^2e^{2\kappa_+b_*} \cosh^2\kappa_+t\right]\,,
\end{equation}
which gives the correct behavior for the entanglement entropy at early times. 

For the early times, no island is formed and we refer to the discussion with no island in Sec.~\ref{sec:noislands}. In this case, the entanglement entropy grows approximately linearly with time as the Hawking radiations which are entangled with the black hole enter into the cutoff surface. 

\subsection{Late times}
At late times, as more and more radiation enters into the cutoff surface, the contribution from the radiation outside the cutoff surface grows. The perpetual linear increase of entropy is what should be expected for the coarse-grained entropy, but for the fine-grained entropy such behavior is prohibited by the unitarity. A simple argument was given by D. Page, that at the early stage when the subsystem is substantially smaller than the total system, the entanglement entropy can be approximated by the thermal entropy of the subsystem \cite{Page:1993wv,Page:1993df}. Exploiting this argument, we should expect the fine-grained entropy goes linearly at the beginning of the radiation. At the later stage of the evaporation, we can apply this argument again to with the small subsystem replaced by the black hole. Then, we expect a linear decrease of the black hole fine-grained entropy. In this section, we will calculate the fine-grained entropy explicitly and see if implementing one island will resolve the information issue at late times of the radiation.
 
We will proceed by assuming $t_a, t_b\gg b > r_+$. We first extremize the generalized entropy Eq.~\eqref{islands} with respect to time $t_a$. The time-dependent component of the generalized entropy is given by
\beq S_{\mathrm{gen}} \supset \frac{c}{3} \log\left\{ \cosh\kappa_+t_a\cosh\kappa_+t_b  \cdot\frac{\cosh\kappa_+(r_*(a)-r_*(b))-\cosh\kappa_+(t_a-t_b)}
 {\cosh\kappa_+(r_*(a)-r_*(b)) + \cosh\kappa_+(t_a+t_b)}\right\} \ . 
\eeq
Employing the following approximations, 
\beq \cosh\kappa_+t_{a,b}\simeq \frac{1}{2}e^{\kappa_+t_{a,b}} \,,\eeq
and \beq \cosh\kappa_+(t_a+t_b) \gg \cosh\kappa_+(r_*(a)-r_*(b))\,, \eeq
and we have the following time-dependent entropy expression,


\beq  S_{\mathrm{time}} \simeq \frac{c}{3} \log\left\{\cosh\kappa_+(r_*(a)-r_*(b))-\cosh\kappa_+(t_a-t_b)\right\} \, \footnote{In principle we should also include the absolute value sign inside the logarithm, but since the points a and b are on the same Cauchy slice we can ignore the negative value issue.}. 
\eeq
We can readily observe that the maximal value of the $S_{\rm time}$ is obtained when $t_a=t_b$ under the approximations made. When substituting $t_a=t_b=t$, we notice that the explicit time dependence disappears from this equation, which is the early sign that the entropy is bounded! 
 
Now we invoke the above and the following approximations,
 \beq \cosh\left(\kappa_+(r_*(a)-r_*(b))\right)\simeq \frac{1}{2} e^{ \kappa_+(r_*(b)-r_*(a))}\,,\eeq 
 and rewrite it in terms of {\rn} coordinate variables,
  \begin{align}
    \cosh\kappa_+(r_\ast(a)-r_\ast(b)) \simeq & \frac{1}{2}e^{\kappa_+(b-a)}\left|\frac{b-r_+}{a-r_+}\right|^{\frac{\kappa_+r_+^2}{r_+-r_-}}\left|\frac{b-r_-}{a-r_-}\right|^{\frac{-\kappa_+r_-^2}{r_+-r_-}}\notag\\
    = & \frac{1}{2}e^{\kappa_+(b-a)}\left|\frac{b-r_+}{a-r_+}\right|^{\frac{1}{2}}\left|\frac{b-r_-}{a-r_-}\right|^{\frac{-r_-^2}{2r_+^2}}
    \,. \label{coshapprox}
 \end{align} 
We incorporate the above approximations. After some algebra we have the generalized entropy $S_{\rm gen}$ read
\begin{align}
S_{\mathrm{gen}} &= \frac{2\pi a^2}{G_{\rm N}} + \frac{c}{6} \log \left[2^4 f^2(a)f^2(b)e^{2\kappa_+(r_\ast(a)+r_\ast(b))}\cosh^2(\kappa_+t_a)\cosh^2(\kappa_+t_b)\right] 
\notag\\
 &\quad
 + \frac{c}{3} \log \left[\frac{\cosh\left(\kappa_+(r_*(a)-r_*(b))\right)-\cosh\left(\kappa_+(t_a-t_b)\right)}
 {\cosh\left(\kappa_+(r_*(a)-r_*(b))\right) + \cosh\left(\kappa_+(t_a+t_b)\right)}\right]\notag \\
 &\simeq \frac{2\pi a^2}{G_{\rm N}} + \frac{c}{6} \log \left[2^4 f^2(a)f^2(b)e^{2\kappa_+(r_\ast(a)+r_\ast(b))}\cosh^2(\kappa_+t_a)\cosh^2(\kappa_+t_b)\right] 
 \notag\\
 &\quad  + \frac{c}{3} \log \left[ \frac{\frac{1}{2} e^{ \kappa_+(r_*(b)-r_*(a))}-1}
 {\cosh{\kappa_+(r_*(b)-r_*(a))} + \cosh\left(\kappa_+(t_a+t_b)\right)}\right] \notag\\
 & \simeq \frac{2\pi a^2}{G_{\rm N}} + \frac{c}{6} \log \left[f^2(a)f^2(b)\right] + \frac{c}{3} \log \left[ \frac{1-2e^{- \kappa_+(r_*(b)-r_*(a))}}
 {1+ e^{ \kappa_+(r_*(b)-r_*(a)-2t_b)}}\right]+\frac{2c}{3}\kappa_+r_*(b)\notag\\
 & \simeq \frac{2\pi a^2}{G_{\rm N}} + \frac{c}{6} \log \left[f^2(a)f^2(b)\right]   +\frac{2c}{3}\kappa_+r_*(b) \notag\\
 &\quad +\frac{c}{3}\left[-2e^{-\kappa_+(r_*(b)-r_*(a))}-e^{ \kappa_+(r_*(b)-r_*(a)-2t_b)}\right]\notag\\
 & \simeq \frac{2\pi a^2}{G_{\rm N}} + \frac{c}{6} \log \left[ f^2(a)f^2(b)\right]+\frac{2c}{3}\kappa_+r_*(b) \notag \\ 
 &\quad  + \frac{c}{3} \left[-2e^{-\kappa_+(b-a)}\left|\frac{a-r_+}{b-r_+}\right|^{\frac{1}{2}}\left|\frac{a-r_-}{b-r_-}\right|^{\frac{-r_-^2}{r_+^2}}-e^{ \kappa_+(r_*(b)-r_*(a)-2t_b)}\right] \, ,
\end{align}
where in the equation behind the second approximately equal sign we used the approximation \beq 2 \cosh(\kappa_+t_a)\cosh(\kappa_+t_b)\simeq \cosh(\kappa_+(t_a+t_b))\,.\eeq
In the equation behind the third approximately equal sign we assumed the first order expansion in the logarithm, and in last line we applied the approximation Eq.~\eqref{coshapprox}.

We notice that the expression has a weak dependence on the time $t_b$ as indicated from the last term. However, this term is of higher order and its magnitude decays exponentially as time goes on. Therefore, the exact location of the island depends on the time. However, its dependence rapidly dies off in a inversely exponential manner and reaches to the asymptotic value. Here, we consider the later stage of the evaporation where the information bound might be broken and ignore the subdominant terms. One of the remarkable differences between this result and the previous one assuming no island (Eq.~\eqref{noislands}) is the disappearance of the explicit time dependence at late stage in the generalized entropy. This implied that after extermization, we would have an answer that is independent of the time, which suggests a convergent behavior of the entanglement entropy instead of going linearly with time. 


We extremize this result with respect to $a$ and we obtain the location of the extremal island as follows,
\beq a\simeq r_++\left[ \frac{G_N c\cdot e^{\kappa_+(r_+-b)}}{12\pi r_+\sqrt{b-r_+}}\left(\frac{r_+-r_-}{b-r_-}\right)^{-\frac{r_-^2}{r_+^2}}\right]^2\,. \eeq
The boundary of the island locates slightly outside the outer horizon and it covers the inside of the horizon as shown in Fig.~\ref{fig:penroseisland}. The higher order correction to the location of the island is dependent on the location of the cutoff surface. Suppose that we set the cutoff surface close to the horizon $b\rightarrow r_+$ (though strictly speaking the validity of our calculation in this case is questionable), the second order term vanishes and the location of the island is closer to the horizon. Therefore, the higher order terms can be understood partially as the artifact of the arbitrariness of the cutoff surface and the boundary of the island lies within the stretched horizon.

\begin{figure}[t]
\centering
	\includegraphics[width=110mm]{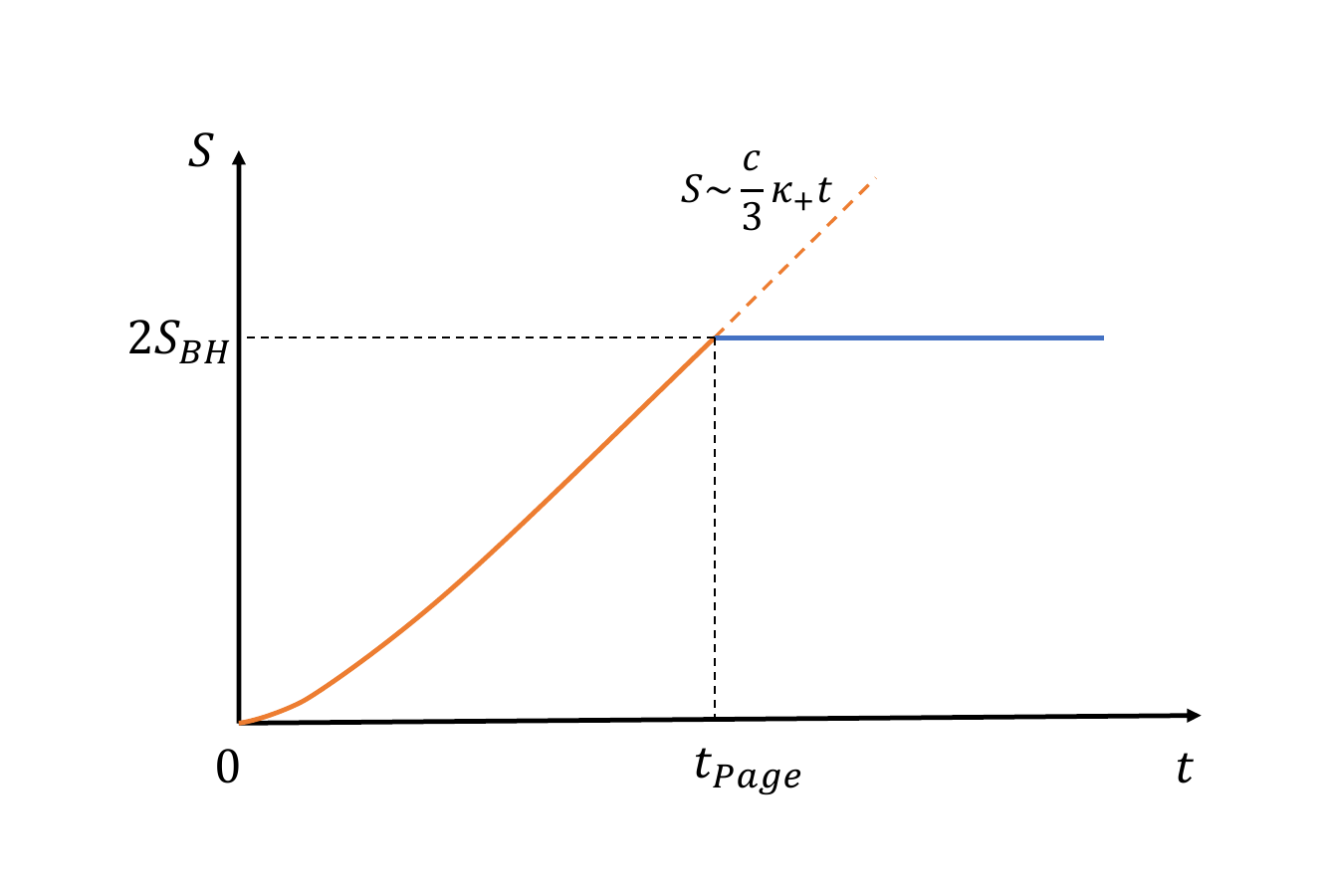}
	\caption{
	The Page curve for the eternal {\rn} black hole with the assumption that the higher order terms in $c \, G_{\rm N}/r_{\rm h}^{2}$ are ignored. The orange dashed line shows the result without islands. The solid line represents the quantitative result with island.}
	\label{fig:Page}
\end{figure}

From the location of the island, we have the entanglement entropy as follows,
\beq S_{\rm entanglement} \simeq \frac{2\pi r_+^2}{G_N}+\frac{c}{3}\log\left[\frac{r_-}{b\cdot\kappa_+^2}\left(\frac{r_-^2}{(r_+-r_-)(b-r_-)}\right)^{\frac{\kappa_+-\kappa_-}{2\kappa_-}}e^{2\kappa_+(b-r_+)}\right]+\ \text{small} \,.\label{entropyisland}
\eeq
The first dominant term of this equation is the Bekenstein-Hawking entropy which naturally comes out from the island construction. The maximal entropy of the radiation is the black hole thermodynamic entropy in the $t\rightarrow\infty$ limit, or equivalently when infinite amount of radiation has been generated by the black hole. Combining the results of the early times and the late times, we have the Page curve which incorporates the early linearly growth of entropy as indicated from Eq.~\eqref{noislands} and the later constant behavior from Eq.~\eqref{entropyisland}. The Page curve is shown in Fig.~\ref{fig:Page}. It should be noted that the higher order terms in $c \, G_{\rm N}/r_{\rm h}^{2}$ are negligible compared to  $t_{\rm Page}$ or $S_{\rm BH}$ and are ignored. This shows the preservation of the information during the evaporation of black holes and resolves the potential information paradox since the entanglement entropy is bounded by the Bekenstein-Hawking entropy of the {\rn} black hole which is finite. This is in contrast to the Hawking's original argument of infinite radiation entropy at the end stage of eternal black holes, which directly violates the unitarity and the information conservation. 

Similarly, the AMPS firewall paradox can be avoided in the {\rn} black hole due to the appearance of the island in the later stage of the black evaporation. The appearance of the islands renders some degrees of freedom of the black hole interior to be inside the entanglement wedge of the radiation. Therefore, not all the degrees of freedom inside the black hole should be counted as the black hole degrees of freedom but only the ones in its entanglement wedge. Therefore, the assumption on the degrees of freedom of black holes which was made implicitly in the AMPS proposal should be released and no firewall near the black hole horizon is expected.

\subsection{Page time and scrambling time}
\label{sec:page}
The time when the entropy of the radiation reaches the maximum is the called the Page time. In an evaporating black hole, it is the time after which the entropy of the radiation starts to decline and around when the black hole is roughly half of its initial mass. For an eternal black hole, its entropy will be a constant after the Page time. 

The Page time for the {\rn} black hole can be inferred from comparing the expressions of the entropy without island [Eq.~\eqref{noislands}] and with island [Eq.~\eqref{entropyisland}]. The two curves cross at approximately the time when the transition happens from the no island configuration to the island configuration, and the entanglement entropy stays at a constant value after that. From this argument, we can calculate the Page time to be as follows,
\beq t_{\rm Page}\sim 6\pi r_+^2/(c\cdot \kappa_+G_N)=\frac{12\pi r_+^4}{c\cdot G_N(r_+-r_-)}=\frac{3}{2\pi c}\frac{S_{\rm BH}}{T_{\rm RN}}\,,
\eeq
here $S_{\rm BH}$ is the Bekenstein-Hawking entropy for the pair of black holes, which is double the entropy of a single black hole. This Page time is comparable to the black hole half life time. Combining the results from Eq.~\eqref{noislands} and Eq.~\eqref{entropyisland}, we obtain the Page curve for the evolution of radiation entropy, or equivalently of the entropy of the black hole as shown in Fig.~\ref{fig:Page}. The entropy of the radiation increases approximately linearly with time in the first stage of the evaporation, during which no island is formed. Around Page time, the island forms near the horizon of the black hole and the entropy stays nearly constant which is twice the thermal entropy of the {\rn} black hole.

According to the Hayden-Preskill protocol, the scrambling time dictates how long the dictionary thrown into a black hole can be decoded from the outgoing Hawking radiation \cite{Hayden:2007cs}. In the language of the entanglement wedge construction, the scrambling time corresponds to the time when the information enters into the island. We assume that the information is retrievable immediately after that through deciphering the radiations. The location of the island is related to the scrambling time in the way such that a light ray sent off from the cutoff surface a scrambling time ago intersects with the boundary of the island. In the case of eternal black holes, the location of the island is fixed near the horizon. Therefore, the scrambling time is essentially the boundary time that the null rays takes to reach the island. 

Suppose we send a message from the cutoff surface at $r=b$ to the black hole, the time it takes to reach the $r=a$ is given as follows,
\beq \Delta t=b-a+\frac{r_-^2}{r_+-r_-}\log\frac{a-r_-}{b-r_-}-\frac{r_+^2}{r_+-r_-}\log\frac{a-r_+}{b-r_+}\,.
\eeq
Given that the island is located at 
\beq a\simeq r_++\left[ \frac{G_N c\cdot e^{\kappa_+(r_+-b)}}{12\pi r_+\sqrt{b-r_+}}\left(\frac{r_+-r_-}{b-r_-}\right)^{-\frac{r_-^2}{r_+^2}}\right]^2\,, \eeq
we can calculate the time it takes for the null rays to enter the island. Once the information is in the entanglement wedge, we assume that the information is retrievable. Therefore, the location of the islands leads to the scrambling time as follows, 
\beq  t_{\rm scramble}\simeq \frac{2r_+^2}{r_+-r_-}\log\frac{r_+^2}{G_N}+{\rm small}\simeq \frac{1}{2\pi T_{\rm RN}}\log S_{\rm BH} \,.\eeq
Under the assumption that the central charge is much less than the degrees of freedom of the black hole $c\ll S_{\rm BH}$, the scrambling time we obtained logarithmically smaller than the life time of the black hole. The leading order is consistent with the result derived in refs.~\cite{Sekino:2008he,Hayden:2007cs}, which is negligible compared to the Page time. This scrambling time from the {\rn} black hole corroborates the argument of the fast-scrambling of information of black holes. 

\section{Discussion}
Though the original black hole information paradox is phrased in the context of evaporating black holes, the paradox and the physics can be equally addressed in the context of eternal black holes. For an evaporating black hole, the amount of radiation is bounded. After the black hole disappears the fine-grained entropy has to be zero at the end of the evaporation process. This is at odds with the thermal Hawking radiation. For an eternal black hole, the amount of radiation is infinite at the ``end stage" of the evaporation and the black hole still has the same mass or charge as the initial conditions. Then the thermal radiation that produces infinite amount of entropy of the radiation is contradictory to unitarity which dictates that the maximal entropy produced by the black hole is the Bekenstein-Hawking entropy. The two versions of the paradox are essentially the same. The AMPS firewall problem can similarly be applied to eternal black hole scenarios, and predict a constant firewall at the horizon \cite{Almheiri:2012rt,Almheiri:2019yqk}. However, this melodrama can be avoided from the argument of ER=EPR \cite{Maldacena:2013xja}. 

In summary, in this study we investigated the information problem of the well-known solutions to Einstein's equation, the 4-dimensional {\rn} spacetime. In the initial period, no island is formed. This is due to the fact that the quantum entanglement entropy for the gravitational system is approximately the minimal of the two components in many circumstances, one is the area of the island and the other is the radiation in its entanglement wedge. At the early stage, not enough radiation has been produced and the contribution to the entanglement entropy mainly comes from the radiation and no island is needed. At the late time stage, the radiation becomes the predominant term and the entanglement is mainly from the area of the island, which lies within the stretched horizon. Using the configuration of the island, we derived the scrambling time that is consistent with that given by Hayden-Preskill protocol and the Page time.

The Penrose diagram in our analysis is for non-extremal black holes. The non-extremal RN black hole and the extremal black hole solutions are not topologically connected and they admit different Penrose diagrams. One should be aware that the extremal or near extremal black holes are unstable and have zero or near zero neutral radiation but vibrant charged radiation. When the charge is much smaller than the mass of the black hole and the Hawking process dominates over the Schwinger process, our analysis is valid. Toy models of eternal extremal and near-extremal black holes without the above consideration are discussed in Ref.~\cite{Almheiri:2019yqk} in which the black holes are assumed to be static and only Hawking process is assumed. 

Moreover, in our construction we only considered the case with zero or one island. In general any patterns of island formation are possible. In our Page curve a sharp turning point at the Page time appears, multiple islands around the Page time will presumably soften the edge of the Page curve. Besides, from the configuration of islands we have answered the question whether black hole information is conserved in the 4-dimensional {\rn} background. However, the possible dynamics describing how the information leaks out into the radiation zone is still lacking. One tentative argument is given by the ER=EPR to explain the information leakage \cite{Maldacena:2013xja}, which can also be suggested from the entanglement wedge, but a concrete mathematical framework incorporating the graviton degrees of freedom at the level of quantum states is yet to be established. 

\acknowledgments
R.L. would like to thank Hongbao Zhang and Yuxuan Liu for useful discussions.

\appendix


\end{document}